# Engineering mass transport properties in oxide ionic and mixed ionic-electronic thin film ceramic conductors for energy applications

*Iñigo Garbayo[1], Federico Baiutti[1], Alex Morata[1] and Albert Tarancón[1,2],\**

[1] Department of Advanced Materials for Energy Applications, Catalonia Institute for Energy Research (IREC), Jardins de les Dones de Negre 1, 08930 Sant Adrià del Besòs, Barcelona, Spain.

[2] ICREA, Passeig Lluís Companys 23, 08010, Barcelona, Spain

\* Corresponding author e-mail: atarancon@irec.cat

**Abstract**

New emerging disciplines such as Nanoionics and Iontronics are dealing with the exploitation of mesoscopic size effects in materials, which become visible (if not predominant) when downsizing the system to the nanoscale. Driven by the worldwide standardisation of thin film deposition techniques, the access to radically different properties than those found in the bulk macroscopic systems can be accomplished. This opens up promising approaches for the development of advanced micro-devices, by taking advantage of the nanostructural deviations found in nanometre-sized, interface-dominated materials compared to the "ideal" relaxed structure of the bulk. A completely new set of functionalities can be explored, with implications in many different fields such as energy conversion and storage, or information technologies. This manuscript reviews the strategies, employed and foreseen, for engineering mass transport properties in thin film ceramics, with the focus in oxide ionic and mixed ionic-electronic conductors and their application in micro power sources.

# 1. Introduction

The upcoming revolution of the Internet of Things (IoT), with an expected market of billions of miniaturised devices (wireless sensor nodes, WSN) installed by 2025, will transform the way in which complex physical systems are understood and interact. [1] Miniaturisation is the meaningful and cost-effective way of expanding the number of WSN of a network, following the "*smart dust*" concept introduced by Prof. Pister in 2001, based on 1 $cm^3$ piconodes. [2,3] Integration of microsensors, communication components and power supply in autonomous small MicroElectroMechanical Systems (MEMS) is the goal, though not yet fully accomplished. The major limitation nowadays is on the energy supply, where relatively large batteries are today the dominant technology (for unwired power solutions). State-of-the-art liquid-based battery capacity remains an unresolved serious problem hindering a number of applications, especially when miniaturised nodes are involved or embedded energy is required. Space constraints in WSN demand smaller batteries which feature an even more limited energy capacity. As a result, in many cases, primary batteries can provide energy only for a fraction of the operation life of the electronic device itself (months vs years), which implies regular battery replacement with subsequent important economic and environmental downsides.

In this scenario, thin film ceramics are destined to play a crucial role on the development of high-performing powering devices, eventually able to substitute the low performing state-of-the-art liquid-based batteries. [4] The interest in functional ceramic metal oxides, with either pure ionic or mixed ionic electronic conduction (MIEC), is huge for their application in more efficient, more compact solid state electrochemical powering devices, e.g. all solid state batteries (Li-conductors) [5] or solid oxide fuel cells (oxide-conductors) [6] as well as for their implementation in other different types of electrochemical devices such as gas sensors. [7]

Traditionally, the typical low mobility associated to the relatively big moving species in oxide ionic conductors hindered a broad implementation of these materials. In many cases, their applicability was limited to high operating temperatures and thus the possibility of integration in portable electronics was restricted. In this context, size reduction and the use of sub-micrometre thin films with a lower associated resistance meant a first breakthrough on the way to ceramics integration in lab-on-a-chip systems. [8,9] Also, integration in silicon was accomplished by using state-of-the-art physical vapour deposition techniques (e.g. sputtering, pulsed laser deposition or atomic layer deposition) and new powering devices such as micro solid oxide fuel cells (μSOFC) are being developed with the benefits of (i.) a reduced operating temperature and (ii.) a fast start-up with low power consumption. [10] However, still insufficient ion conductivity or, in other words, too high operating temperature (T > 400 °C) in the functional ceramic films has been hindering their broad implementation, and a new step forward is now needed.

New emerging disciplines such as Nanoionics and Iontronics, dealing with the design and control of interface-related phenomena in fast ionic conductors, have recently opened new routes for the use of functional thin film ceramics in optimised powering systems working at reduced temperatures. [11–23] The high presence of interfaces in thin films entails important size effects that might be used for the control and the tuning of materials' properties. [24,25] Different phenomena such as lattice strain, accumulation of defects or local non-stoichiometry, can highly impact the electrochemical properties of ceramic interface-dominated thin films and can have profound implications on the realisation of new devices. Generally, Nanoionics deals with the study of the local compositional and microstructural modifications taking place at an interface. Meanwhile, main concept of Iontronics is the use of electrochemical double layer (EDL) capacitances occurring between an ionic and an electronic conductor for inducing very high density carrier concentration channels. Also, interface-dominated materials are being studied for enhancing other mass transport phenomena, e.g. the Oxygen

Reduction Reaction (ORR) capabilities of ceramic electrodes for SOFC systems. [4,26,27] All in all, a *nanoionics revolution* is foreseen, similar to the one driven by nanoelectronics few decades ago.

In this feature review, we give an overview of the different aspects accounting for the control and the development of thin film interface-dominated ceramic materials, for their final use as oxide ionic or mixed ionic-electronic conductors in portable powering microdevices. In a first section, we describe the growth methods, the possible architectures and the controlling phenomena encountered in this type of materials. Recent results showing important enhancements in the transport properties of ceramic materials are discussed. The second part of the review focuses on the implementation of nanoionics/iontronics ground-breaking concepts into real devices, still a rather unexplored field mainly because of the difficult integration of genuine interface phenomena in technologically relevant substrates.

## 2. Interface-dominated thin film ceramic materials

In nanocrystalline materials, the proportion of interfaces is maximised and pronounced size effects may become predominant over the bulk behaviour (mesoscopic regime). Interfaces are defined here as two-dimensional defects irrevocably formed during fabrication; at an interface, the bulk crystalline symmetry is broken and a redistribution of species should be expected. [4] In the core of the interface, a modified (often off-equilibrium) structure may form. [18] Here, we analyse interface-dominated materials from two points of view: their architecture, viz. homo or heterointerfaces, vertically or horizontally aligned, and the phenomena by which their properties can be tuned, namely lattice strain, structural defects, local composition, space-charge and, ultimately, light.

*2.1 Interface architectures in thin film ceramic materials*

While fundamental studies on ionic and MIEC systems have traditionally dealt with 3D macrostructures such as ceramic pellets, it becomes clear that the implementation of such materials on real micro devices should rely on thin film technology. Here, great control over the boundary conditions and the geometry can be achieved; moreover, the alternation of the different layers (e.g. electrodes and electrolyte for SOFC application) can be obtained at the nanometre-scale, this way achieving a high interface density. Most importantly, thin film technology is compatible with micromachining techniques, as we detail in section 3.

Different strategies can be implemented for the control of interfaces in thin film ceramics, giving rise to distinct interface architectures. First, it is obvious that the choice of the deposition method will greatly influence the properties of the films, and that an extremely precise and versatile technique is needed for obtaining high quality and controllable interfaces. Second, tuning the deposition conditions may determine whether the films are grown in a single crystal fashion or polycrystalline, with clear implications on the number and type of interfaces generated. If growing single crystal, one can still envision further engineering by playing with the substrate, or by fabricating composite films in the form of multilayers or nanocolumns. In the case of polycrystalline films, the orientation of the grains can also play a major role on the local changes happening at the interfaces.

2.1.1. The growth of interface-dominated thin films

Multiple deposition techniques have been developed in the last decades for the growth of oxide thin film structures, classified in two main categories: Physical Vapour Deposition (PVD) (such as sputter deposition, Pulsed Laser Deposition (PLD) and Molecular Beam Epitaxy (MBE)) and Chemical Vapour Deposition (CVD) (mainly Atomic Layer Deposition (ALD) and Chemical Vapour Deposition (CVD)) methods. Among such techniques, sputtering and CVD have been the most commonly employed at the industrial level given the possibility of covering large substrate areas. However, such methods seem to suffer from intrinsic limitations (such as a poor stoichiometric control, presence of secondary phase impurities and low reactivity or volatility of metal-organic precursors) which make their implementation for the fabrication of nanometre-controlled complex oxides heterostructures cumbersome. Several successful CVD and sputtering growth of complex oxides are however worth of mention (see e.g. [28,29] and references therein).

In very recent times, particular interest seems to have arisen from the possibility of exploring ALD for the realisation of oxide thin film nanostructures. ALD is a layer-by-layer technique in which the substrate is cyclically exposed to chemical precursors which, unlike CVD, are introduced in the chamber one at a time, Figure 1a. Although the amount of material which can be deposited at each cycle is self-limited (which makes the technique intrinsically slow), large areas can be covered in a very conformal, pinhole-free, fashion. [30,31] Traditionally, thin films of binary oxides have been fabricated by ALD; however, much effort has been recently put on the fabrication of multi-component oxides with high crystallinity and minimised impurity content. Some examples mainly related to perovskites can be found in literature. [32,33]

To date however, the fabrication of interface-dominated complex oxide films for research purposes has mainly relied on the PLD and the oxide MBE methods, due to their high versatility in the choice of the growing compounds, a great control over the layer thickness (potentially down to the atomic layer level), a superior interface quality and phase purity. PLD and MBE are particularly well suited for thin film fabrication in a layer-by-layer fashion (i.e., the full coverage of the film surface is achieved prior to the formation of a new layer). [34] Such a growth mode allows for the minimisation of the density of extended defects and for the formation of smooth interfaces. Extensive descriptions over thin film epitaxy and growth modes in general can be found e.g. in references [29,35], while dedicated reviews of PLD and MBE techniques can be found elsewhere; [36,37] here, some account is given.

PLD and MBE working principles rely on the evaporation of material atoms or clusters and subsequent deposition on the surface of a heated substrate. In the case of PLD, Figure 1b, material is evaporated in the form of a plasma, which is formed by the bombardment of a ceramic pellet (or single crystal) with the desired stoichiometry by a pulsed UV laser. In case of MBE, Figure 1c, an atomic flux is generated by thermal evaporation or sublimation of metals. Depending on the material vapour pressure, evaporation in an oxide MBE system can be carried out by means of resistive heater effusion cells (Knudsen cells) or by an electron-beam evaporator. The deposition of oxides typically requires the delivery of an oxidising gas (molecular oxygen, oxygen plasma or ozone) during the film growth. This is a particularly crucial point in the case of oxide MBE in which, due to the low kinetic energy of the evaporated species, a near-ballistic path is required inside the chamber (i.e. $p < 10^{-4}$ Torr). For this purpose, purified ozone is the most common gas of choice. [38]

The standard procedures and the wide commercial widespread of the components, which characterize the PLD method, make its implementation for the fabrication of complex oxide multilayers relatively easy and cost-effective. The versatility of the PLD method allows for the growth of thin films with completely different microstructures, from dense to porous, and from polycrystalline to single crystals. However, the complex phenomena which are related to the plasma formation, to its interaction with the background gas and to the film phase nucleation and growth are still object of extensive investigation. [39,40] It should be mentioned in particular that substantial deviations of the cationic stoichiometry, due e.g. to scattering, to the different volatility of the species or to material re-evaporation, may occur, [41–44] potentially having important effects on the final film properties. [45,46] An accurate control of the deposition conditions is therefore needed.

Oxide MBE systems have emerged as a very powerful tool for multicomponent oxide fabrication in recent years. [37,47,48] Although mainly devoted to the fabrication of oxides for electronics, such a technique has shown its potential also in relation to solid state ionics as demonstrated e.g. by Sata et al. [49] and Azad et al. [50] The main advantages of such a technique with respect to the PLD are related to the presence of single-element fluxes, to the minimised kinetic energy of the impinging species (< 0.1 eV) and to the slow kinetics of the process, which ensure a great control over the growth process. Particularly interesting is the atomic-layer-by-layer (ALL) method in which, owing to an accurate control system of the source shutter timing, the chemical composition of each atomic layer can be defined allowing for the fabrication of multilayers having the highest quality and single-layer engineered structures. [37,51–53] The main drawback of the MBE system is, together with a high system complexity, related to the challenge of maintaining the desired cationic stoichiometry. [54] For this, a number of in-situ monitoring tools such as quartz crystal microbalance (QCM), real-time absorption spectroscopy (AAS), [55] electron impact emission spectroscopy (EIES), [56] and especially reflection high-energy electron diffraction (RHEED) are commonly employed. [57–59] However, it should be mentioned here that both PLD and MBE methods require the use of expensive deposition systems which are characterised by high level of complexity and whose process scale-up is far from being accomplished (see also Section 3). For these reasons, much effort should be put for the implementation of such techniques for real device fabrication, i.e. for large area deposition.

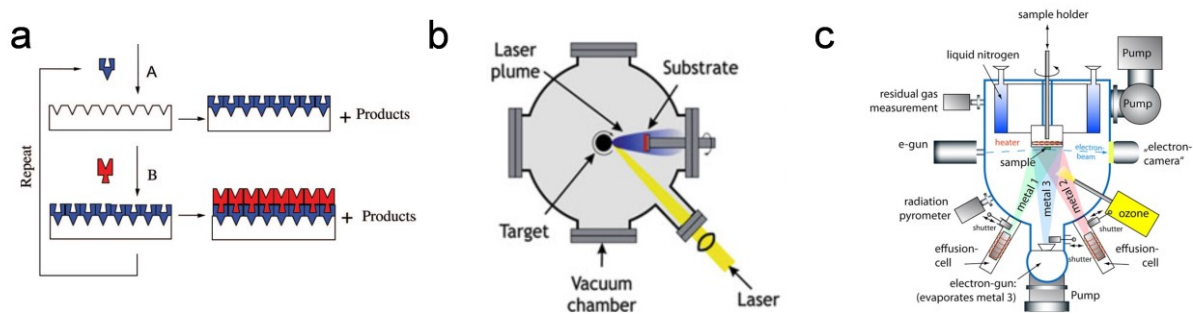

*Figure 1. Working principles of different techniques for the deposition of high quality complex oxides: (a) Atomic Layer Deposition (ALD) (reprinted with permission from S.M. George, Chem. Rev. 110 (2010) 111–131. Copyright 2010 American Chemical Society), [30] (b) Pulsed Laser Deposition (PLD) (reprinted by permission from RightsLink Permissions Springer Customer Service Centre GmbH: Springer Nature, Laser Ablation and Thin Film Deposition by C.W. Schneider, T. Lippert Copyright 2010) [60] and (c) Molecular Beam Epitaxy (MBE) (reprinted from F. Baiutti et al., Beilstein J. Nanotechnol. 5 (2014) 596–602). [61]*

2.1.2. Types of interfaces in thin films

A qualitative classification of interfaces can be made on the basis of the constituting phases, being either one single material (homointerfaces, i.e. grain boundaries), or two different materials (heterointerfaces), see Figure 2. [14] Grain boundaries (GBs), which are defined by the misorientation of two crystallographic planes from two adjacent grains in polycrystalline materials, [62,63] probably represent the most typical example of how conduction properties in a ceramic can be intrinsically altered by microstructural modifications. In thin films, it is possible to obtain grains of < 50 nm in diameter, i.e. the GB density is significantly high. [64] Therefore, profiting from the modified defect chemistry of GBs represents a highly powerful strategy for tuning the functionalities and for the development of engineered materials. Together with this, great interest has arisen on the study of heterointerfaces as a controlled way to tune the properties of ceramic materials. This approach entails the growth of epitaxial layers and multilayers.

In this section, a general overview of the possible type of interfaces and related phenomena is given. A more specific description on interface effects is reported in section 2.2.

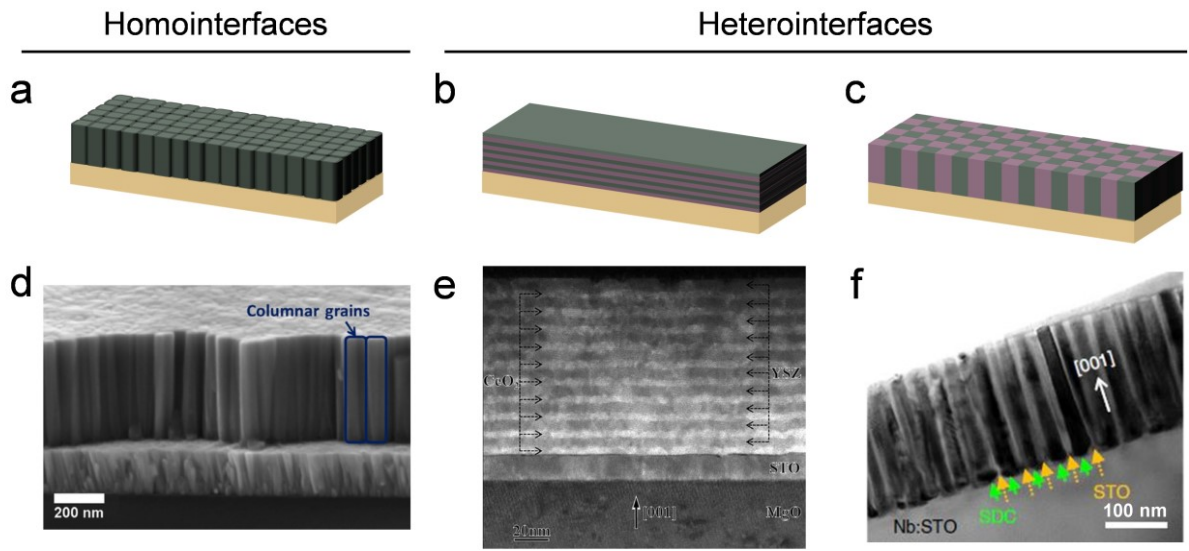

*Figure 2. Types of interfaces in ceramic oxide films: Homointerfaces, i.e. grain boundary-dominated materials (a,d) [65] and heterointerfaces, viz. multilayers (b,e) [66] and vertically aligned nanocomposite structures (VANs) (c,f). [67] (d) is reprinted from Metal Oxide-Based Thin Film Structures, Elsevier, F. Chiabrera, I. Garbayo, A. Tarancón, Nanoionics and interfaces for energy and information technologies, 409-439,, Copyright 2018, with permission from Elsevier. (e) is reprinted with permission from D. Pergolesi, E. Fabbri, S.N. Cook, V. Roddatis, E. Traversa, J.A. Kilner, ACS Nano. 6 (2012). Copyright 2012 American Chemical Society. (f) is reprinted from S.M. Yang, S. Lee, J. Jian, W. Zhang, P. Lu, Q. Jia, H. Wang, T.W. Noh, S. V Kalinin, J.L. MacManus-Driscoll, Strongly enhanced oxygen ion transport through samarium-doped $CeO_2$ nanopillars in nanocomposite films, Nat. Commun. 6 (2015) 8588. Copyright 2015.*

*Homointerfaces: Grain boundary-dominated materials*

It is well known that grain boundaries, due to their high degree of disorder and accumulation of defects, greatly affect the concentration (and mobility) of ion species, thus influencing the ionic conduction. [12,68] Traditionally, grain boundaries in bulky ceramic systems were seen as a drawback. In classical thick, randomly-oriented polycrystalline systems, any device configuration forces to cross multiple of these highly disordered regions, which always adds an extra resistance to the ion movement and worsens the material´s performance. However, the use of thin film deposition processes has lately changed this paradigm, as it allows to tune the grain size, the level of disorder and the orientation of the interfaces. Instead of seeing the grain boundaries as an obstacle to be crossed by the moving species, structures can now be fashioned in such a way that conduction happens *along* the boundary (Figure 2a,d).

Depending on the degree of misorientation θ, grain boundaries can be classified in low- and high-angle GBs, with profound implications in the atomic arrangement. In general, a low angle grain boundary can more easily accommodate the planes' misorientation by forming arrays of dislocations. Using Frank´s equation, [69] one can predict the periodicity of the rearrangement as a function of the misorientation angle θ, viz. the higher the angle the lower the distance between dislocations. High misorientation angles lead to defect overlapping, generating more complex structures. Numerous studies, focused on resolving the local atomic changes happening in grain boundaries, appeared in the last years with the aim to understand and ultimately tune the relationship between the GB atomic structure and electro-chemo-mechanical properties of the material. In general, it can be affirmed that, in the vast majority of oxide materials, grain boundary cores are characterised by a reduced formation enthalpy for oxygen vacancies. As a consequence, vacancy accumulation occurs and a positively charged GB core is formed. [70,71] It is noteworthy, though, that a few examples of negatively charged core have been also reported, i.e. dislocations in polycrystalline $TiO_2$ [72] and $Y_2Zr_2O_7$. [73] In those cases, the presence of negative charges in the dislocation cores has been ascribed to the accumulation of cation vacancies. [72,74] In any case, charge compensation mechanisms lead to additional local alterations in the vicinities of the GB core, which include the formation of space charge regions and/or cationic rearrangements (see section 2.2).

In oxygen ion conductors such as yttria-stabilised zirconia (YSZ) or gadolinia-doped ceria (CGO), the accumulation of oxygen vacancies in the GBs has been widely reported and important implications in the oxygen conduction properties have been proven. [75–84] On the one side, by analysing single grain boundaries in YSZ bicrystals, it was evident that a drop on the oxygen ion diffusion *across* the grain boundary occurs. [78] Such a finding is in good agreement with the typical behaviour found in polycrystalline samples, where diffusion across GBs is considered as a high resistance path. [85] However, enhanced oxygen exchange kinetics has been highlighted along GBs. [26,79] Lee et al. measured YSZ polycrystalline films with varying grain sizes (i.e. different GB density at the free surface), proving that increasing the GB density lowers the electrode impedance and increases the exchange current density. [26] Similar results were found for ceria-based thin films. [84] Importantly, this points out the relevance of grain boundary engineering for enhancing the ORR rates in solid oxide fuel cells.

In more recent times, the relevance of grain boundaries in perovskite-related materials has also been addressed. Special attention has been put on the enhanced mixed ionic-electronic properties of strontium-doped lanthanum manganites (LSM). [27,86–88] Two independent studies by Saranya et al. [27,88] and Navickas et al. [86] reported fast oxygen conduction properties through grain boundaries of LSM. In both works, it was shown how grain boundaries improve the mixed ionic electronic character of a mainly electronic conductor such as LSM, thus opening up the possibility for LSM to be employed as an electrode for solid oxide fuel cell (SOFC) systems. Both oxygen surface exchange and oxygen diffusivity are significantly enhanced in LSM GBs as compared to the bulk. It is expected that these results can be extended to other perovskite-related materials with potential application as cathodes in μSOFC systems.

*Heterostructures: multilayers and vertically aligned nanocomposites*

At a heterointerface, contact is established between two phases having nominally different chemical and structural composition. Heterostructure technology possesses a long-standing tradition in the field of semiconductors which has been later applied, in more recent times, in the field of oxides for electronics. [89–93] The purposeful investigation of thin film heterostructures in solid state ionics has seen his breakthrough with the experiments of Sata et al. on $CaF_2$/$BaF_2$ heterostructures, in which an increase in the parallel ionic ($F^-$) conductivity of several orders of magnitude was achieved by increasing the interface density. [49] This was rationalised in the light of the space-charge theory describing $F^-$ accumulation at the interface, see section 2.2.3. [21]

Unlike homointerfaces, in which the interface is defined by the breaking of the crystal symmetry, it is important here to notice that coherent heterostructures can be realised in which the crystal continuity is maintained at the phase contact (epitaxial layers, see Figure 2b,e). In a "perfectly" epitaxial system, a pseudomorphic growth takes place, *i.e.* the in-plane lattice parameter of the film perfectly couples to the lattice parameter of the substrate. More often, semicoherent interfaces are obtained (see also section 2.2.1). Although epitaxial heterolayers have been often put in relation to strain effects, it has been shown that such phase contact represent a prosperous playground for a number of effects related e.g. to cationic intermixing and charge redistribution, as we describe in section 2.2.

In the last few years, a new architecture for thin film composites has strongly emerged as a possible alternative to the "classical" multilayer structures in which the interfaces are parallel to the free surface. Such an approach, which has been proposed by MacManus-Driscoll and co-workers, relies on the fabrication of vertically aligned nanocomposite structures (VANs), in which the two constituting materials are alternated in the form nanocolumns, whose typical width is in the range of tens of nanometres (Figure 2c,f). [94] Such structures result from a self-assembling PLD process (starting from a two-phase target) once conditions related to the choice of materials (i.e. high chemical compatibility) and to a good lattice-match with the substrate (at least for one of the two phases) are met. The advantages of such a technique are related e.g. to the possibility of being employed in applications in which an access to interfaces in an out-of-plane geometry is required. VANs structure have been found to exhibit surprising effects related e.g. to fast ionic conductivity, [67] magnetic effects [95] and resistive switching. [96] Extensive reviews over the main results have been published recently. [97,98] It should be noted here that, owing to the high coherence of the interfaces and to the particular geometry, very high strain states (up to $\varepsilon$ = 4-5 %), which can't be achieved on "regular" multilayer systems, can be obtained. [99]

*2.2. Interface nanoengineering: The controlling phenomena*

2.2.1. Strain effect

Very often, the contact between two differently oriented grains or two different materials leads to the formation of a strain field given by the deformation of the crystal lattice with respect to the bulk material, Figure 3a. A typical model system is represented by epitaxial interfaces in thin film multilayers. In such systems, the in-plane elastic deformation of the film (epitaxial strain) can be easily calculated:

$$\varepsilon = \frac{a_s - a_b}{a_b} \quad (1)$$

with $a_s$ being the substrate lattice parameter and $a_b$ being the bulk (unstrained) lattice parameter of the film. The corresponding out-of-plane strain $\varepsilon_{zz}$ is related to $\varepsilon$ via the Poisson ratio $\nu$:

$$\varepsilon_{zz} = \frac{2\nu\varepsilon}{(\nu - 1)} \quad (2)$$

A certain elastic energy can be normally stored in the film. However, epitaxial strain is released by the formation of extended defects (typically dislocations or differently oriented domains) if the thickness of the film is increased over a critical value. Several analytical expression for the calculation of the critical thickness $t_c$ have been proposed (see e.g. Frank-van der Merve model [100] and Matthews-Blakeslee model [101]). Typical values for $t_c$ are around ≈ 10-20 nm for $\varepsilon$ ≈ 0.5-1 %, but this is greatly dependent on the growth kinetics. For larger strain values, the appearance of defects typically occurs already at the very initial stages of the growth.

It is very important to notice, however, that only very narrow conditions result in an epitaxial growth. Most commonly, thin film structures develop through the formation of islands (Volmer-Weber growth) or in a mixed fashion, layer-plus-island or Stranski-Krastanov, resulting in films having disordered or columnar domain structure (the latter being typical of PLD-grown systems). [64] Therefore, other sources of local deformation should be taken into account together with epitaxial strain and the resulting stress state in a film should be considered as a sum of competing processes occurring during the growth. For instance, the formation of a grain boundary due to the coalescence of two isolated islands is associated with the development of a tensile stress, [102–104] while migration of materials from the surface to the grain boundary or impingement of energetic particles (e.g. during the PLD growth) give rise to compressive strain. [105,106] Particularly relevant in this context is the case in which coherent interfaces are obtained in columnar multilayered structures: Here the lattice mismatch between the constituting phases is able to generate a new strain field at each interface irrespective of the total film thickness. [107,108]

Several computational studies have been dedicated to the prediction of the effects of strain on ionic conductivity. Generally, a decrease of the migration enthalpy is expected as a consequence of a moderate tensile strain state, as pointed out by Yildiz et al. [109] and De Souza et al. [110] which predicted an enhancement of diffusivity of ≈ $10^3$ times for 4 % strained YSZ and $CeO_2$, respectively. Tarancón et al. reported on a possible oxygen conductivity increase of up to two orders of magnitude for $\varepsilon$ < 4 % (tensile strain). [111] Interestingly, the presence of a critical strain upon which conductivity is expected to decrease was highlighted.

Extensive experimental investigations regarding strain effects on ionic conductivity have been carried out on YSZ/fluorite-related systems. Korte et al. observed a slightly increased ionic conductivity and tracer diffusivity in the YSZ/$Y_2O_3$ system, characterised by semicoherent interfaces (YSZ under tensile strain). [112,113] Similar results were also obtained by Li et al. on multilayers of YSZ and $Gd_2Zr_2O_7$ grown on $Al_2O_3$ (0001). [114] Results regarding the YSZ/ceria system are quite contradictory: Azad et al. [50] and Sanna et al. [115] studied heterostructures of doped ceria and YSZ, highlighting a sizeable increase in the ionic conductivity with increasing number of interfaces. However later works by Pergolesi et al. on related systems showed a negligible effect of interfacial conductivity on transport properties, as retrieved by electrochemical and tracer diffusion experiment. [66] Moreover, complementary investigations (including HR-TEM spectroscopy experiments) by the same author on $Ce_{0.85}Sm_{0.15}O_{2-\delta}$/YSZ heterostructures pointed out that an insulating layer (namely, cubic bixbyite $Ce_2O_3$) forms at the ceria side of the interface, leading to an overall decreased interlayer conductivity. [116]

A very controversial example is given by $SrTiO_3$/YSZ heterostructures, related to which Garcia-Barriocanal et al. reported up to eight orders of magnitude ionic conductivity increase and a decrease of the activation energy of 0.5 eV. This was ascribed to the presence of interface tensile strain up to 7 %. [117] However, such results have been greatly debated by further studies on similar PLD structures, which highlighted a non-negligible contribution of electronic conductivity stemming from $SrTiO_3$ (STO). [118,119]

Overall, the effects of strain on ionic conducting structures are quite elusive. Among the influencing factors, one should consider the very poor structural compatibility between the typically employed substrates and the films, as well as differences in the preparation conditions. This leads to a difficult control over the presence and the density of extended defects (dislocations and grain boundaries), which possess a different chemical environment and peculiar functionalities. Moreover, even in the case of "ideal" interfaces, different effects including anionic and cationic redistribution (interdiffusion or space-charge effects) may come to the fore. [120] As highlighted by Korte et al. [121], pure strain effects should be restricted to systems with extrinsic conductors having very short Debye lengths.

2.2.2. Dislocations

Dedicated studies have been carried out in relation to concentration and mobility effects induced by line defects. Static lattice calculations by De Souza et al. [122] and atomistic simulations by Yildiz and co-workers [123] on the model system STO indicate oxygen vacancies segregation in dislocation cores due to a reduced defect formation energy (about 2 eV with respect to the bulk), see Figure 3b. However, due to their low mobility this does not correspond to high oxygen diffusion; rather, this leads to the formation of negatively charged vacancy depletion space-charge zones surrounding such positively charged dislocations. Such a scenario was confirmed by tracer diffusion experiments by the same authors. [122] More recent computational studies by Sun et al. on doped $CeO_2$ also predict deviation of the charged defect concentrations around dislocations due to steric effects, leading to a decreased oxide ion diffusion. [124] Finally, as already mentioned earlier, works by Maier group on $TiO_2$ [74] and on ionic conducting $Y_2Zr_2O_7$ [73] suggest that cationic vacancy segregation in the dislocation core may occur, leading to the formation of a negatively charged core and to an enrichment of oxygen vacancies within the adjacent space-charge regions. In these cases, a positive effect of dislocations on oxygen ionic conductivity was observed.

2.2.3. Space charge

Space-charge effects play a crucial role for the modification of the material functionalities by confinement. [19] The basic concept is that any interruption of the crystal continuity (e.g. dislocations, grain boundary, interfaces) inherently brings-in an electrical charge leading to the formation of a space-charge potential (e.g. due to concentration of vacancies, defect segregation, cationic and anionic inter-diffusion, Figure 3c). [70,92,125,126] As a consequence, a local redistribution of all the mobile (ionic and electronic) defects is expected in conditions of thermodynamic equilibrium (i.e. constancy of the electrochemical potential for each defect specie) within the so-called space-charge region across the interface, as described by Poisson-Boltzmann equation: [11]

$$\frac{\partial^2 \varphi}{\partial x^2} = -\frac{e}{\varepsilon_0 \varepsilon_r} \sum_j z_j c_{j,\infty} \exp\left(-\frac{z_j e \Delta\varphi(x)}{kT}\right) \quad (3)$$

From this equation, one can estimate the spatial variation of the concentration $c$ for each defect species at a distance $x$ from the interface. Here $\varphi$ and $z$ represent the space-charge potential and the defect charge number, respectively. $e$ refers to the elementary charge, $k$ to the Boltzmann constant, $\varepsilon_0$ is the permittivity of free space, $\varepsilon_r$ the relative permittivity of the material and $T$ the temperature in Kelvin. Equation (3) can be solved analytically under specific boundary conditions (and in the limit of dilute defect concentrations), namely (i.) in the presence of two defects with opposite charge number (Gouy-Chapman situation) and (ii.) when the concentration of the majority defect is considered flat and frozen-in, while the compensating counter-defect is depleted (Mott-Schottky situation). [127] It is important to notice that the space-charge width is expected to decrease by increasing the dopant content in the bulk, being as high as tenths of nanometres in slightly doped materials, [128] and 1-2 nm in the case of heavily doped systems. Space-charge solutions for interacting defects have been recently developed by Mebane and co-workers. [129]

Space-charge situations have been intensively investigated in a number of ionic and mixed ionic-electronic conductors (extensive reviews can be found in ref. [65,130,131]). In most cases of technological relevance, such as perovskites and fluorite-based materials, [128,132–134] grain boundary cores have been found to be positively charged, i.e. space-charge regions are blocking layers for oxygen vacancy conduction due to concentration depletion effects. In case of mesoscopic systems, in which the Debye length becomes comparable to the grain size, huge variations in the values and in the type of conductivity has been found. For example, Lupetin et al. observed an increase in the n-type conductivity of up to 3 orders of magnitude and a decrease of the p-type conductivity (by the same extent) and of the oxygen vacancy conductivity (of up to 6 orders of magnitude) in $SrTiO_3$ upon grain size reduction (down to 30 nm); [128] qualitatively similar effects have been found for nanocrystalline ceria. [135] So far, successful attempts of tuning the space-charge potential and to obtain vacancy accumulation regions in the proximity of interfaces have been quite limited. [136–139]

2.2.4. Local non-stoichiometry

As mentioned earlier in this paper, the presence of positive defects, mainly oxygen vacancies, in the interfacial core is the most common situation in ceramic interfaces, and is linked to the high strain values and less spatial constrains present in the interface. [123] Note also here the few examples of negatively charged core reported by Maier and co-workers, where an accumulation of cation vacancies in the interface core occurs. [72–74] Anyway, for all the cases, a redistribution of the concentration of other elements is usually observed, as a partial compensation mechanism of the core charge generated, see Figure 3d. [14] The extent and expanse of the local non-stoichiometry is not trivial, and depends on the characteristics of the charged core (interface potential), type of interface (misorientation, concentration of defects…), the defect chemistry of the material(s), and finally the mobility of the species. It is therefore linked with the formation of space charge regions (see previous section), in which the redistribution of species strongly depends on the (dopant) cation mobility. The width and significance of the space charge can be modified, for instance, by changing the doping level of a material. Increasing the cation doping, the space charge width usually decreases due to a higher capability of charge screening with the accumulated dopant. Above a certain doping value, more complex compensation mechanisms can however appear, giving rise to dopant-defect interactions and complex distributions of species along the space charge. [129]

Dopant segregation and cation redistribution has been commonly found in grain boundaries of fluorite oxide conductors like YSZ and CGO, together with the oxygen non-stoichiometry of the GB core. [140,141] Frechero et al. reported the segregation of Y to the grain boundaries in YSZ accompanying an oxygen depletion, confined to a very small length scale of ca. 0.5 nm (i.e. basically in the GB core), Figure 3d. [76] Such cationic redistribution might even induce phase transitions in the vicinity of YSZ grain boundaries, as suggested by Feng et al. [77] Similarly, Li et al. found segregation of Gd in CGO, forming small nanodomains at the grain boundaries. [142] Sun et al. associated the dopant segregation in CGO to the high strain fields formed around dislocations, which detrimentally affect the ionic conductivity in those regions. [124] Direct evidences of dopant segregation are supported by several computational studies, see e.g. the work by Yoshima and Oyama on different YSZ grain boundaries. [143] Importantly, a correlation between the dopant segregation energy and the concentration of oxygen vacancies was established, being favourable the formation of defect complexes at the grain boundary level.

Local stoichiometry changes have been observed as well in heterointerfaces. Here, particular relevance should be given to cationic inter-diffusion, which may occur across interfaces due to the strong cationic concentration gradients and to the high lateral mobility of the atoms during preparation. [144] Such an effect has been very intensively investigated in the case of perovskite and derived structures (electronic and mixed ionic-electronic conductors) in which, although spatially limited to few nanometres, they have been shown to be potentially responsible for deep effects in the interface functionalities (e.g. by providing unintentional cationic doping). [145–149] Aidhy and Weber [14] reported a computational analysis of the interplay between lattice strain generated at $CeO_2$ interfaces and dopant segregation. They simulated the effect of compressively strained $CeO_2$-$ZrO_2$ and tensile strained $CeO_2$-$ThO_2$ interfaces on the segregation energy of different elements. Preferential dopant segregation towards the heterointerface was found for tensile strains, whereas dopants tend to stay in the bulk for compressive interfacial strains.

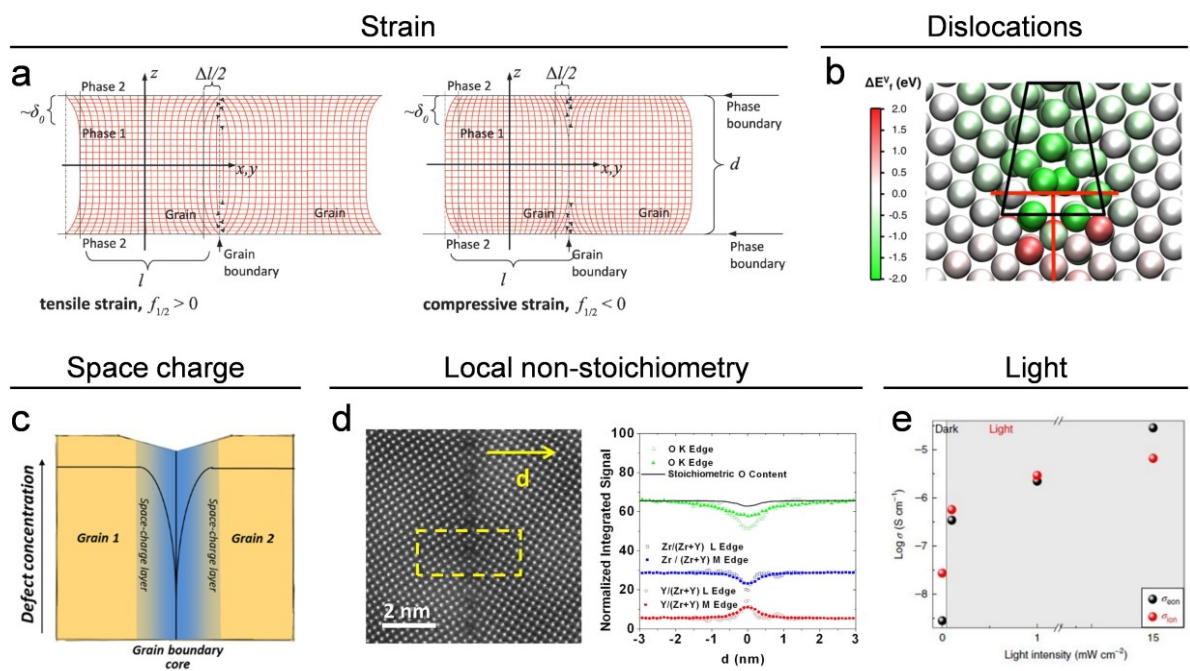

*Figure 3. The controlling phenomena for engineering mass transport properties in ceramic thin films. (a) Crystalline deformations caused by tensile and compressive strain (reprinted from C. Korte, J. Keppner, A. Peters, N. Schichtel, H. Aydin, J. Janek, Phys. Chem. Chem. Phys. 16 (2014) 24575–24591 - Published by the PCCP Owner Societies); [121] (b) the effect of dislocations: map of the change in formation energy of oxygen vacancies in $SrTiO_3$ dislocations (Reprinted with permission from D. Marrocchelli, L. Sun, B. Yildiz, Dislocations in $SrTiO_3$: Easy to reduce but not so fast for oxygen transport, J. Am. Chem. Soc. 137 (2015) 4735–4748. Copyright 2015 American Chemical Society); [123] (c) the concept of space charge formation exemplified in a grain boundary; (d) local non stoichiometry changes, as observed in a YSZ grain boundary (reprinted from M. Frechero, Paving the way to nanoionics: atomic origin of barriers for ionic transport through interfaces, Sci. Rep. (2015) 1–9. Copyright 2015); [76] (e) light effects in the electrochemical properties of ceramics: increase in ionic and electronic conductivities of a metal halide perovskite electrolyte by illumination (reprinted by permission from RightsLink Permissions Springer Customer Service Centre GmbH: Springer Nature, Nature Materials, G.Y. Kim, A. Senocrate, T.-Y. Yang, G. Gregori, M. Grätzel, J. Maier, Large tunable*

*photoeffect on ion conduction in halide perovskites and implications for photodecomposition, Nat. Mater. 17 (2018) 445–449, copyright 2018). [150]*

*2.3. Light-driven phenomena*

There is an increasing interest on the exploitation of light and light-driven phenomena in ionic and mixed ionic electronic conductors for enhancing the properties of electrodes in energy systems. In this direction, it is particularly relevant the pioneering work of Merkle et al. [151] where the authors proved that the rate of oxygen exchange is enhanced by a hundredfold under UV light illumination for a model system such as the wide bandgap Fe-doped $SrTiO_3$ electro-ceramic oxide. This result reinforces the idea that the governing species for oxygen incorporation in SOFC cathodes are not the majority charge carriers (oxide ions and holes in the state-of-the-art p-type semiconductor materials) but the minority charge carriers, i.e. electrons, as suggested by Jung and Tuller in a recent work. [152] Contrary to the previous understanding of the oxygen reduction reaction (ORR), the excited electrons would control the kinetics of the oxygen surface exchange, i.e. control the rate determining step at low temperatures, and, therefore, the study of mechanisms to promote electrons to the conduction band should be prioritised to the enhancement of the concentration of majority carriers or their mobility. Despite these inspiring initial results, the field remains almost unexplored and work is needed in order to explore this effect in other cathode materials and in real device configurations.

Another potential benefit of illumination in ceramics could be related to the increase of the ionic conductivity. Until very recently, only a few studies indirectly suggested a possible effect of light in the ionic conduction of ceramics. However, Maier and co-workers [150] have now reported that illumination can increase by orders of magnitude not only the electronic conductivity, but also the ionic conductivity in methylammonium lead iodide, an archetypal metal halide perovskite photovoltaic material, see Figure 3e. A mechanism for the ionic conduction improvement based on a reaction of the generated electron-holes to form iodine vacancies, and the subsequent generation of interstitial neutral iodine, was proposed. The exploitation of this phenomena may lead to a whole new class of photoactivated solid state electrochemical devices based on light-induced conduction changes. [153]

Finally, two recent works from Fleig´s group deserve to be highlighted here. [154,155] First, Brunauer et al. showed that light can be efficiently used for pumping oxygen through thin film perovskite-based cells, when stacking a photovoltaic cell and an electrochemical cell, all based in ceramics. [154] This way, the proof-of-concept of a solar to fuel conversion device by using light-driven ceramic electrochemical cells was established. In a second study, Walch et al. showed the appearance of a Nernst type voltage in solid oxide electrochemical cells based on $SrTiO_3$ and YSZ, under UV illumination. [155] A time-dependent voltage was generated in the cell, induced by changes on the oxygen stoichiometry (δ) of $SrTiO_{3-\delta}$ caused by the UV-driven oxygen incorporation. This can lead to the realisation of solid oxide photoelectrochemical cells, or "light-charged oxygen batteries", in which the capacity is defined by the amount of oxygen incorporated under illumination. [156] Eventually, this work indicates the importance of having available excited electrons in the oxygen incorporation process.

**3. Implementation of interface-dominated materials in novel solid state energy devices**

Efforts have been put in recent years in order to allow for the practical implementation of real device systems taking advantage of the Nanoionics concepts. At the present stage, one can recognize at least two limiting factors: (i.) the small area capabilities of the above-described mainstream growth methods, i.e. PLD and MBE (Figure 1b,c) and (ii.) the use of substrates based on ceramic single crystals, which have a low technological relevance due to their poor functionalities and to the incompatibility with conventional microfabrication processes. Therefore, in order to have a real impact in the MEMS industry, the integration of Nanoionics concepts should rely on large area deposition of ceramics on reliable substrates compatible with mainstream microtechnology. The use of silicon as a substrate combined with scalable thin film deposition techniques would be highly valuable by ensuring (i.) the required manufacturability and cost effectiveness; (ii.) the miniaturisation needed to integrate solid state power sources with other micro/nanodevices; (iii.) the possibility of fabricating dense architectures to achieve high specific energy and power and (iv.) a potential seamless integration of multiple components in systems. In this direction, increasing efforts are underway for the integration of functional nano-engineered films into silicon technology in order to implement in real devices the fascinating phenomena discovered at the interfaces between single crystal substrates such as $SrTiO_3$ or $LaAlO_3$ and different semiconductors. It is important to highlight here the ability shown by some research groups [157–159] to implement oxide buffer layers (most typically YSZ or $SrTiO_3$) in order to fabricate epitaxial functional films on conventional Si substrates as well. The use of other sacrificial layer buffers (such as $Sr_2Al_2O_6$) is being explored to obtain free-standing structures of high-quality films able to be transferred from the original substrate to an arbitrary support. [160]

Added to that, as mentioned before the use of a technologically relevant substrate such as silicon should be combined with a further development of batch production deposition systems in which homogeneous conditions (process temperature, growing rate…) are ensured over larger areas. Examples of large area PLD systems are currently available at several research institutes, e.g. University of Twente and IREC Barcelona [161,162], while other strategies like multi-plume implementation and cassette-to-cassette or roll-to-roll loading have been already implemented for positioning this technique in industrial scenarios. [163,164] In this direction, systems specially designed for the deposition of $YBa_2Cu_3O_{7-x}$ (YBCO) capable to produce layers between 1 or 2 μm in 10 $m^2$ substrates in a 10h process have been also developed. [161] Moreover, in the case of ALD, the main challenge nowadays is on finding ways to increase the materials' deposition rates. In this regard, the development of the so-called spatial ALD (SALD), [165] in which precursors are introduced in the chamber separated in space instead of in time, permits to substantially increase the deposition speed, with the additional benefit of not requiring high vacuum, which often makes the processing more complicated and expensive to scale up.

In this section, we briefly review novel solid state energy microdevices currently under development, and we give hints on how Nanoionics may help to improve these and other promising new energy devices based on nanoengineered interface-dominated ceramic films.

*3.1. Micro Solid Oxide Fuel Cells*

Pioneering works on the fabrication of silicon-based powering micro devices such as μSOFCs [166–168] or all solid state microbatteries (out of the scope of this article but reviewed somewhere else [9,169,170]) represent an important breakthrough in the integration of functional ceramic films into real devices. The development of μSOFCs experienced great progress a decade ago, thanks to the downscaling of functional ceramic materials to thin films and to their integration in silicon technology, Figure 4a. [6,171] By reducing the thickness of the typical ceramic electrolyte materials used in traditional SOFC systems (mainly YSZ and CGO), it became possible to reduce the internal resistance and, thus, to operate at much lower temperatures. [172] In addition to that, the use of mainstream silicon technology allowed for the fabrication of low thermal mass ceramic free-standing membranes accessible from both sides (out-of-plane geometry), which represents a prerequisite for the realisation of devices having fast and low power consuming start-up processes. [166] Single cells based on YSZ free-standing electrolyte membranes and symmetric metallic electrodes were tested at temperatures as low as 350 °C, [173,174] thus significantly reducing the typical operating temperature of SOFCs (T > 750 °C), and power outputs above 1 Wcm$^{-2}$ were obtained below 500 °C. [175,176]

Despite the promising results, the development of the technology has suffered a deceleration in the last years, mainly because of the combination of two interconnected factors: First, the state-of-the-art materials which are commonly employed as electrolyte are capable to operate down to ca. 450 °C, where the internal resistance is still low enough not to limit the performance of the cell. [172] Second, at such reduced temperatures, the only available active electrodes nowadays are metallic-based; [8] however, it was shown that the use of metallic electrodes at T > 400 °C causes fast degradation and cell performances drop within a few hours of operation. [167] A new breakthrough is therefore needed to overcome these limitations, and the use of Nanoionics concepts is expected to play a key role. Reducing even more the operating temperature of the cells would allow an easier system encapsulation and would facilitate their integration with other CMOS-based components (usually limited to maximum temperatures of 400 °C). It becomes obvious that electrolytes with higher oxygen ion conductivity are needed. Although one strategy could rely on the substitution of YSZ by other ceramic materials, [177,178] technological limitations due to the need of adapting the microfabrication processes are foreseen. Taking advantage of the Nanoionics concepts possesses a very high potential owing to the superior performances. Examples in this sense, related e.g. to conductivity enhancement via strain or space-charge effects have been commented in this paper. Notably, taking into account the particular geometrical configuration of µSOFC, viz. vertical ion movement crossing the thin film electrolyte, the use of columnar structures (e.g. VANs) in which oxide ion conductivity is enhanced through the vertical heterointerfaces is a promising approach. [67]

In addition, substituting the metallic electrodes by ceramic ones would help to ensure the long term operation of the cells, eliminating the detrimental effect of the metallic film degradation. [167,168] However, state-of-the-art thin film ceramic electrodes are restricted to operation at a relatively high temperature range (T > 700 °C) by their limited activity. [168,179] In order to reduce it, strategies for enhancing the ORR rates in these electrodes are needed. For example, as recently reported by Saranya et al. [27,88] and Navickas et al., [86] both the oxygen incorporation and oxygen diffusion are enhanced by orders of magnitude in $La_{0.8}Sr_{0.2}MnO_{3+\delta}$ (LSM) and $La_{0.8}Sr_{0.2}(Mn_{1-x}Co_x)_{0.85}O_{3\pm\delta}$ (LSMC) grain boundaries as compared to the bulk. Combined with the intentional growth of functional films with vertically aligned grains where GBs represent fast diffusion pathways, this strategy can significantly help to reduce the operating temperature of µSOFC cathodes. An almost unexplored, yet enthralling approach for improving the electrodes activity at reduced temperatures could be the photoexcitation, in ceramic materials in which minority charge carriers (e$^-$) are the rate limiting factor (see section 2.3).

### 3.2. Ion-gated thermoelectrics

Charge carrier concentration and mobility have a major impact on the performance of thermoelectric devices. [180] Up to now, the most common way to increase the carrier concentration has been chemical doping. However, it is known that under certain circumstances the intercalation of dopants can have a deleterious effect in the performance of the materials, since the formation of defects and dopant segregation can also produce a decrease of the electrical conductivity. This effect is particularly relevant in nanostructured materials. Moreover, the doping approach also implies a limitation in the achieved carrier concentration. In order to overcome this difficulty, a second route based on Nanoionics effects can be used, consisting of the use of an electric field applied through a gate material (gating). This procedure has proven to give excellent results, providing carrier concentrations orders of magnitude higher than the maximum achievable by chemical doping. [181–183] Extremely large capacities can be obtained if an electrolyte is used as a gate instead of the thin film high-permittivity insulators which are commonly employed as gates for CMOS transistors in microelectronic industry. In this situation, when a voltage is applied, the ions in the electrolyte redistribute forming the so-called electrochemical double layers (EDLs) at the interface with the semiconductor below. A very large interface capacitance is then generated due to the extremely short distances between the ionic and electronic charges, which is independent on the electrolyte dimensions. Summed to the thermoelectric material properties tuning, ionic gating can be used for the fabrication of thermoelectric transistors, similar to the traditional electronic transistors that are the core of the silicon microfabrication technology. This provides excellent opportunities for new applications which involve the monitoring of low temperature differences requiring low consumption scenarios. Some examples are smart sensors or human-organic electronic interfaces (like functional dressings and fabrics for medical monitoring).

As an example of an interesting application of ion gating in thermoelectrics, Takayanagi et al. [184] proposed a new paradigm for a device in which the thermoelectric properties of both p- and n-legs of a thermocouple are simultaneously controlled by the application of a potential in the common gate. A voltage was applied between the positive $Cu_2O$ electrodes and the negative ZnO electrodes, in order to accumulate holes on the surface of $Cu_2O$ and electrons on the surface of ZnO. As a result, the output power of the device was increased to more than 10 times under certain conditions.

A second example of application uses an electrolyte with high Seebeck coefficient to naturally generate a potential difference in the gate when exposed to a heat source. In the work of Zhao et al., [185] an external heat-gated organic transistor was developed consisting of an electrolyte-gated transistor and an ionic thermoelectric supercapacitor (ITESC), Figure 4b. When the ionic thermoelectric leg is subjected to a temperature gradient ΔT, it produces a high open voltage by the Soret-effect that gates the transistor. Hence, the temperature difference induces a variation in the drain current. Providing an operating voltage of the electrolyte-gated transistor in the same order of the variation of the thermal voltage of the ITESC, the authors were capable to tune the transistor output current in more than two orders of magnitude when applying temperature differences.

*3.3 Solid oxide photoelectrochemical cells*

The development of solid oxide photoelectrochemical cells (SOPECs) relies on the combination of photovoltaics and reversible fuel/electrolysis cells. Such a thermo-regenerative energy system (TRES) was intensively studied in the space age; [186,187] however, up to date the fabrication of SOPECs was exclusively based on two separate cell systems, and only single cells were liquid-based. [188,189] On the way to system miniaturisation, it becomes clear that neither of the latter approaches is ideal, since they both present volumetric constrains and high voltage drops are expected.

Only very recently, first concepts of all-solid-state operating systems were reported by Fleig´s group in Vienna, see Figure 4c. [154,155] As conceptually proven by Fleig and co-workers, directly converting radiant energy into chemical energy within a simple thin film ceramics-based device is possible, by combining an all-solid-oxide photovoltaic (PV) cell and a reversible solid oxide electrochemical (EC) cell in a single, multilayer device. In such a system, efficiency will rely on maximising the interaction between light and the functional ceramic material, for which nanoengineering and Nanoionics are expected to play a key role.

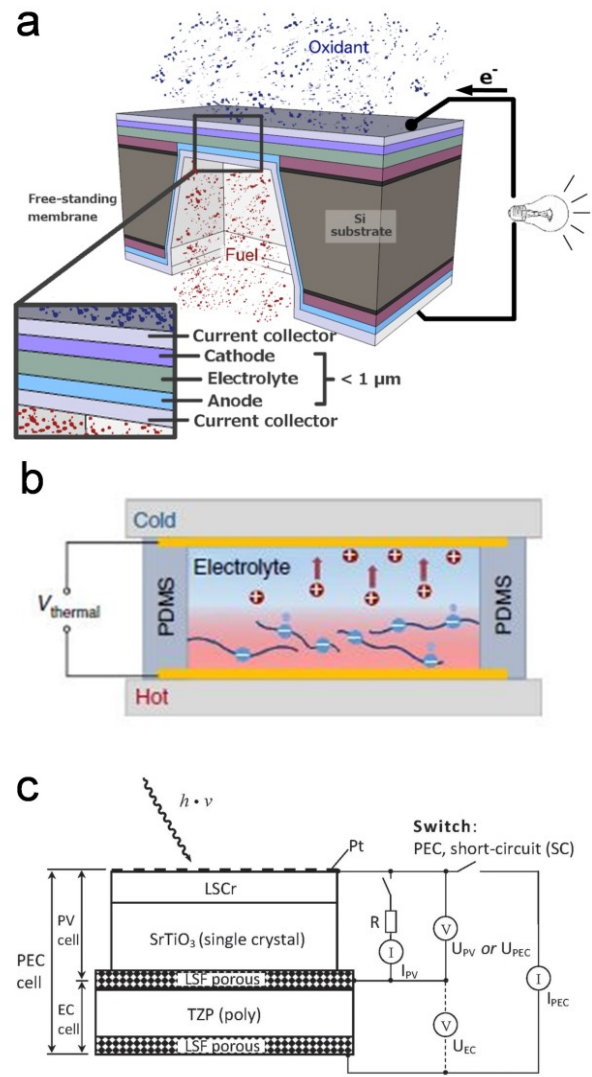

*Figure 4. Novel solid state energy microdevices for the implementation of interface-dominated oxide ceramic thin film materials: (a) micro solid oxide fuel cells (reprinted from I. Garbayo, Integration of thin film based micro solid oxide fuel cells in silicon technology, Universitat de Barcelona, 2013), [190] (b) ionic thermoelectric supercapacitors (ITESC) (reprinted from D. Zhao, S. Fabiano, M. Berggren, X. Crispin, Ionic thermoelectric gating organic transistors, Nat. Commun. 8 (2017) 14214. Copyright 2017) [185] and (c) Solid oxide photo electrochemical (PEC) cell (reprinted with permission of John Wiley and Sons, from G.C. Brunauer, B. Rotter, G. Walch, E. Esmaeili, A.K. Opitz, K. Ponweiser, J. Summhammer, J. Fleig, UV-light-driven oxygen pumping in a high-temperature solid oxide photoelectrochemical cell, Adv. Funct. Mater. 26 (2016) 120–128. Copyright 2015). [154]*

**4. Conclusions and prospects**

A breakthrough in micro-energy technologies is required to cover the increasing demand for embedded, personal or local use of power. Due to their miniaturisation capabilities, solid state energy devices are promising candidates to play a major role in this new era. However, improving their performance while downscaling their size can only be achieved by looking for disruptive concepts in materials capabilities. In this sense, emerging disciplines such as Nanoionics and Iontronics, which explore the complex interplay of electronic and ionic properties at the nanoscale, could enable this micro-energy revolution similarly to what nanoelectronics did for computing and processing. Necessarily, this revolution would rely on the possibility of implementing interface-dominated materials in real devices. Therefore, after decades of mastering thin film techniques for depositing advanced ceramics with enhanced properties, especially at the interface level, focused efforts should be dedicated to develop large-area thin film deposition techniques and to their compatibility with mainstream microfabrication technologies. In order to achieve a high technology readiness level, a proper transfer of the recently discovered functionalities of many oxides, which were proven in model systems, into technologically relevant substrates like silicon should be achieved. This extension to more realistic systems will involve the use of easily implementable interface-dominated microstructures such as polycrystalline nanomaterials or vertically aligned nanocomposite structures (VANs). Polycrystalline materials and VANs are dominated by grain boundary interfaces, which clearly affect the mass transport and catalytic properties of oxides. However, these interface effects started to be explored few years ago by the Solid State Ionics community and a universal description of local modifications occurring at the grain boundary level is still far. Grain boundary defect chemistry is a complex phenomenon that depends on multiple factors, including the material local stoichiometry, doping level, strain and/or degree of misorientation of the GBs. Therefore, it seems clear that further understanding of the interplaying phenomena happening at the grain boundaries is still needed in order to reach full control of the relevant processes and thus gain engineering possibilities. Due to this lack of knowledge and the technological gap explained above, the number of devices currently existing based on these novel concepts is still small. Despite this, recent advances, as detailed in this feature review, and pioneering works on developing novel power sources, indicate the way to convert interface-dominated ceramics based on Nanoionics and Iontronics concepts into the main player of a new generation of highly performing micro-energy devices.

**Acknowledgments**

The research was supported by the Generalitat de Catalunya-AGAUR (2017 SGR 1421), the CERCA Programme and the European Regional Development Funds (ERDF, "FEDER Programa Competitivitat de Catalunya 2007-2013"). This project has received funding from the European Research Council (ERC) under the European Union's Horizon 2020 research and innovation programme (ULTRASOFC, Grant Agreement number: 681146).